# Anisotropy of Thermal Conductivity of Free-Standing Reduced Graphene Oxide Films Annealed at High Temperature

J.D. Renteria[1], S. Ramirez[1], H. Malekpour[1], B. Alonso[2], A. Centeno[2], A. Zurutuza[2], A.I. Cocemasov[1,3], D.L. Nika[1,3] and A.A. Balandin[1,*]

[1]Nano-Device Laboratory (NDL) and Phonon Optimized Engineered Materials (POEM) Center, University of California – Riverside, Riverside, California 92521 USA

[2]Graphenea Inc., 1 Broadway, Cambridge, Massachusetts 02142 USA

[3]Department of Physics and Engineering, Moldova State University, Chisinau, MD-2009, Republic of Moldova

## Abstract

We investigated thermal conductivity of free-standing reduced graphene oxide films subjected to a high-temperature treatment of up to 1000°C. It was found that the high-temperature annealing dramatically increased the in-plane thermal conductivity, $K$, of the films from ~3 W/mK to ~61 W/mK at room temperature. The cross-plane thermal conductivity, $K_{\perp}$, revealed an interesting opposite trend of decreasing to a very small value of ~0.09 W/mK in the reduced graphene oxide films annealed at 1000°C. The obtained films demonstrated an exceptionally strong anisotropy of the thermal conductivity, $K/K_{\perp} \sim 675$, which is substantially larger even than in the high-quality graphite. The electrical resistivity of the annealed films reduced to 1 $\Omega/\square$ – 19 $\Omega/\square$. The observed modifications of the in-plane and cross-plane thermal conductivity components resulting in an unusual $K/K_{\perp}$ anisotropy were explained theoretically. The theoretical analysis suggests that $K$ can reach as high as ~500 W/mK with the increase in the $sp^2$ domain size and further reduction of the oxygen content. The strongly anisotropic heat conduction properties of these films can be useful for applications in thermal management.

[*] Corresponding author (AAB): balandin@ee.ucr.edu ; web: http://ndl.ee.ucr.edu/

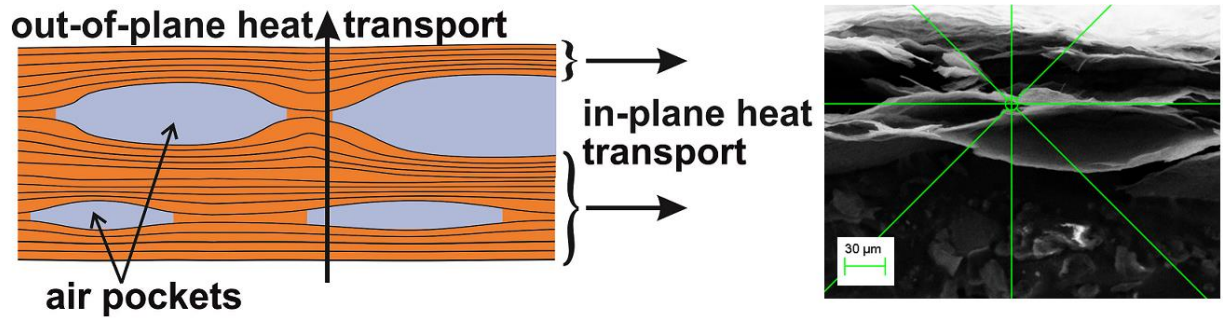

**Contents Image**

Thermal management of modern electronics requires thin films with highly anisotropic thermal conductivity, where the in-plane thermal conductivity, $K$, is substantially larger than the cross-plane thermal conductivity $K_\perp$ [1]. The function of thermal pads or coating layers with this property is to conduct heat away from the hot spots in the in-plane direction while protecting electronic components underneath them from heating. High-quality bulk graphite is an anisotropic heat conductor with the thermal conductivity along the basal planes of $K \approx 2000$ W/mK and the cross-plane thermal conductivity $K_\perp \approx 20$ W/mK at room temperature (RT) [2-3]. Despite its large $K$ and $K/K_\perp \sim 100$, unprocessed graphite cannot be used for thermal management because it does not meet industry requirements, e.g. flexibility. The commercial thermal pads based on composites with chemically processed graphite have lower $K$ and smaller $K/K_\perp$ ratios. These considerations create motivations for the search of new materials with high $K/K_\perp$ that can be used for thermal management, including the removal of excess heat (high $K$) and shielding from excess heat (lower $K_\perp$).

The discovery that graphene has extremely high in-plane thermal conductivity, which can exceed that of the basal planes of graphite [4-8], stimulated a surge in experimental and theoretical studies of heat conduction in graphene and other two-dimensional (2-D) materials [9-14]. Graphene and few-layer graphene (FLG) layers were proposed for thermal management applications as fillers in the thermal interface materials [15-17] or flexible heat spreaders for local hot-spot cooling in electronics and optoelectronics [18-21]. One of the practically feasible ways to industry-scale applications of graphene-based materials for thermal management lies in the reduction of graphene oxide (GO). A well-established Hummers method or its modifications [22-23] allows one to mass produce GO from natural graphite. Thin films on the basis of GO have been shown to have excellent mechanical properties [24-25]. However, GO reveals very low thermal conductivity $K$=0.5 W/mK – 1 W/mK [26-28] at RT. The reduction of GO films to obtain the reduced graphene oxide (rGO) films via conventional chemical or thermal techniques does not necessarily results in increased $K$ or $K/K_\perp$ ratio owing to residue impurities, defects and disorder [8].

In this Letter we report our finding that annealing of the free-standing GO films at temperature, $T \sim 1000^\circ$C, results in substantially increased in-plane thermal conductivity and, simultaneously, reduced cross-plane thermal conductivity. The anisotropy of the thermal conductivity attains a record-high value of $K/K_\perp$ = 675 making the flexible rGO films attractive for thermal management applications. Moreover, the exposure of rGO films to $1000^\circ$C temperature treatment reduces their electrical resistivity to 1 $\Omega/\square$ – 19 $\Omega/\square$, which is the smallest reported to date for such materials. The theoretical considerations suggest that the in-plane thermal conductivity and $K/K_\perp$ can be increased farther via the control of the sp$^2$ grain size and oxygen residue. The rest of the Letter is organized as follows. We first outline the preparation and structural characterization of the free-standing GO and rGO films, describe the measurements of the thermal conductivity by two different techniques, and then offer a theoretical model that captures the main features of the heat transport in such materials.

The samples investigated in this study included the reference free-standing GO films and the films annealed at different temperatures: $300^\circ$C, $600^\circ$C and $1000^\circ$C. The temperature treatment results in reduction of GO films to obtain the free-standing rGO films. The samples were prepared by casting a GO dispersion into a mold and then drying, first at room temperature and then at 60°C in a vacuum oven overnight. For the thermal treatments, the samples where placed in a tube furnace and heated up in a N$_2$ atmosphere. The residence time for each temperature was of 60 minutes. Additional details of the sample preparation and characterization are provided in the *Methods* section.

Figure 1 shows the top (a-c) and cross-sectional (b-d) scanning electron microscopy (SEM) images of the free-standing films. The representative samples are the reference GO film and rGO film annealed at $600^\circ$C. The cross-sectional SEM images reveal layered structure where individual continuous sp$^2$ layers become larger. The interlayer distance between sp$^2$ atomic planes does not change substantially after the thermal treatment in line with previous reports that used XRD analysis [29]. At the same time, the "air pockets" develop between the layers as a result of oxygen and carbon dioxide release. The "air pockets" impede strongly the cross-plane thermal transport while not seriously affecting the in-plane thermal conduction. The surface of the high-temperature annealed films becomes corrugated due to the "air pockets" formation and possible contraction. The morphological changes observed in SEM data are corroborated with the thickness data. The average thickness $H$=40 $\mu$m of the reference GO film increased to $H$=170 $\mu$m in the rGO film annealed at 1000°C. The *true* mass density, $\rho_0$, measured after compacting the material into the special sample container to avoid the air bubbles was 1.87 g/cm$^3$ in the reference GO film. It did not change after the high temperature treatment. The apparent mass density, which includes "air pockets" changed approximately proportionally to the increased thickness (see Methods for details).

[Figure 1 (a-d): SEM]

The chemical composition and morphology of the samples before and after annealing has been studied using X-ray photoelectron spectroscopy (XPS). All samples exhibited carbon (C), oxygen (O), nitrogen (N), and sulfur (S) at varying concentrations. The XPS spectrum range relevant to the present study is shown in Figure 2 (a-b) (see *Supplemental Information* for full spectra). The main peaks observed at ~284.6 eV, 286.8 eV and 288.1 eV correspond to $sp^2$ and $sp^3$ C, single bonded carbon – oxygen (C-O), and double-bonded carbon – oxygen (C=O). Annealing at 600°C results in a significant reduction of both C-O and C=O bonds followed by almost complete extinction of such bonds after 1000°C treatment (see Figure 2 (a) and (b)). The expulsion of O by thermal treatment leads to the concentration of C exceeding 90% after 600°C annealing. The energy difference between C $sp^2$ and C $sp^3$ peaks is rather small, which complicates interpretation. However, the reported data for rGO are in agreement that C $sp^2$ peak energy is in a range 284.1 – 285 eV while C $sp^3$ peak energy in the range 284.9 – 286 eV [30-35]. The measured peak in rGO film annealed at T=1000°C is around 284.8 eV which suggests that it mostly corresponds to C $sp^2$ bonds. This conclusion is in line with the detailed study of the effects of temperature of reduction of GO [29].

[Figure 2 (a-b): XPS]

We used Raman spectroscopy as another tool to monitor how the thermal treatment changes the structural composition of rGO films (see Figure 3 (a-b)). Raman spectroscopy (Renishaw InVia) was performed in a backscattering configuration under visible (λ= 488 nm) and UV (λ= 325 nm) laser excitations. Details of our Raman experimental procedures have been reported by some of us elsewhere [36-37]. Figure 3 (a) shows the Raman spectra for rGO samples that underwent thermal treatment at 300°C, 600°C, and 1000°C. The peaks at ~1350 cm$^{-1}$ and 1580 cm$^{-1}$ correspond to the D and G peaks, respectively. The 2D band centered around 2700 cm$^{-1}$ and a S3 peak near 2900 cm$^{-1}$ are also present and consistent with literature reports for rGO [38-40]. It is observed that the separation between the D and G peaks becomes more pronounced with the sample thermally treated at 1000°C and that the 2D band and S3 peak are also becoming more prominent. These are indications that the GO films underwent reduction as they received thermal treatment and that the films are moving away from an amorphous state to a more ordered material. This is consistent with our XPS data showing that the C=O and C-O bonds have effectively disappeared from the films and helps to explain our thermal data. It is known that UV Raman is more sensitive to C $sp^3$ and C-H bonds. Figure 3 (b) shows that the I(D)/I(G) intensity ratio decreases substantially in rGO annealed at 1000°C as compared to that annealed at 300°C. This suggests that the amount of C $sp^3$ in the samples treated at 1000°C is small and the O reduction mostly results in $sp^2$ bonds, in line with XPSanalysis.

[Figure 3 (a-b): Raman]

The thermal conductivity of the films was studied using the "laser flash" technique (LFT). In addition, the optothermal Raman measurements [8] were performed to cross-check the thermal conductivity values. The LFT transient method directly measures the thermal diffusivity, $\alpha$, of the material [41]. The thermal conductivity is then determined from the equation $K=\rho\alpha C_p$, where $\rho$ is the mass density and $C_p$ is the specific heat of the sample. The specific heat is determined from the independent measurement using a calorimeter or a separate measurement with the same apparatus using a reference sample of similar thermal properties with known tabulated $C_p$ (e.g. graphite). The cross-plane $\alpha$ was measured using the "laser flash" method in the standard configuration: the film is heated by light illumination from one side and the temperature rise is measured on the opposite side. The in-plane diffusivity measurement requires a special sample holder where the location for the light input on one side of the sample and location for measuring the temperature increase on the other side of the sample are at different lateral positions. This arrangement ensures that the measured increase of the sample temperature on the back side corresponds to the thermal diffusivity in the in-plane direction [20]. Details of the measurements are summarized in *Methods* section. Before performing the measurements with rGO films we calibrated the experimental system on several known materials.

Figure 4 shows the cross-plane component of the thermal conductivity, $K_\perp$, as a function of temperature for rGO annealed at different temperatures and a reference GO film. One notices that $K_\perp$ is small for all samples. The most interesting feature is a drastic reduction of $K_\perp$ after high temperature annealing. We explain it by the restoration of the $sp^2$ bonds within the atomic planes and the action of the "air pockets" (the thermal conductivity of the air is ~0.02 W/mK). The softening of the bonds between the layers may also play a role. Below we provide a more detailed theoretical analysis to explain this effect. The cross-plane thermal conductivity only weakly depends on temperature revealing a small growth as $T$ increases. The latter is expected for the amorphous and disordered materials. The film thickness non-uniformity and residue defects suggest that in the cross-plane direction the material is rather disordered even after high-temperature treatment. The $K_\perp$ value of ~0.09 W/mK is extremely small can be considered at the low bound of the amorphous limit [42]. A few cases when thermal conductivity in the cross-plane direction went below the amorphous limit have been reported in literature [43].

[Figure 4: Cross-plane]

The in-plane thermal conductivity, $K$, is presented as a function of temperature in Figure 5. The results are shown for rGO films annealed at $300^oC$, $600^oC$ and $1000^oC$. One can see that the increased temperature of annealing results in higher $K$ values. The room temperature in-plane thermal conductivity increases from 2.9 W/mK for the reference GO film to 61 W/mK for the rGO film annealed at $1000°C$. The rGO films annealed at $300^oC$ and $600^oC$ show increased in-plane thermal conductivity and weak temperature dependence. The slightly increasing $K$ with $T$ for $300^oC$ and $600^oC$ annealed samples suggests that the phonon thermal transport is still limited by disorder. The thermal conductivity of the rGO film annealed at $1000^oC$ reveals decreasing $K$ with $T$ indicative of the onset of the Umklapp-scattering limited phonon transport. The rGO films treated at this high temperature start to behave more like crystalline materials although with a very large concentration of defects. In crystalline materials the phonon thermal conductivity decreases as $1/T$ due to the Umklapp phonon scattering [8]. The increasing $K$ with annealing temperature can be explained by the enlargement of $sp^2$ grains and reduction in phonon scattering on O and other impurities. This interpretation is supported by the XPS data.

To cross-check the thermal conductivity values obtained from the "laser flash" technique we also conducted the optothermal Raman measurements for the $1000^oC$ annealed films. The optothermal technique was initially developed for the measurement of the thermal conductivity of suspended graphene samples [4-6, 8] and later extended to macroscopic suspended films [21]. The temperature rise in response to the laser heating of the samples was extracted from the G and D peak positions. The experimental details are given in the *Supplementary Information*. The obtained value of the thermal conductivity averaged between G and D peak data was consistent with the "laser flash" data within 7% experimental uncertainty.

[Figure 5: In-plane]

It is known that the thermal transport in graphite, graphene and their derivatives is dominated by acoustic phonons [8, 12]. However, it is interesting to compare the changes in the in-plane thermal conductivity with those in electrical conductivity because both the phonon and electron transport can be affected by the defects and structural disorder in carbon materials. We have measured the electrical sheet resistance using the van der Pauw technique. The details of the measurement procedures are provided in the *Methods* section. The main finding was that the annealing at high temperature results in the decrease of the resistivity from 0.5 MΩ/□ range to ~1 – 19 Ω/□ for the rGO film subjected to annealing at T=1000°C. The data for the samples treated at different temperatures are summarized in Table I. The data reported previously for rGO are also provided for comparison in Table II. The increase of the electrical conductivity of rGO films can be explained by increasing $sp^2$ phase as in other carbon derivatives. It is also in line with previous studies of electrical conductivity of rGO, which correlated it with the $sp^2$ (C-C)/$sp^3$ (C-O, O-C-O) peak ratio in XPS spectra [34].

**Table I:** Electrical Resistance of Thermally Treated rGO Films

| Sample | Sheet Resistance |
|---|---|
| GO | 0.514 ± 0.236 MΩ/sq. |
| rGO (T=300°C) | 27.0 Ω/□ ± 17 Ω/ sq. |
| rGO (T=600°C) | 2.01 Ω/□ ± 1.6 Ω/ sq. |
| rGO (T=1000°C) | 2.13 Ω/□ ± 1.1 Ω/ sq. |

**Table II:** Comparison of Electrical Resistance Data for rGO Films

| rGO Sample: Reduction Method | Sheet Resistance | Reference |
|---|---|---|
| Thermal reduction and annealing at 300°C – 1000°C | 1 – 19 Ω/sq | This work |
| Hydrogen and thermal treatment followed by CVD | ~14 kΩ/sq | [44] |
| Hydroiodic acid (HI) | ~840 Ω/sq | [45] |
| Hydrazine vapor and thermal treatment | 100 - 1000 Ω/sq | [46] |
| Thermochemical nanolithography | 18 – 9100 Ω/sq | [47] |
| Hydrogen reduction | 18 Ω/sq | [48] |

For our rGO samples, $T \approx 1000$°C can be considered as a useful reference annealing temperature at which O reduction and $sp^2$ bond restoration leads to a drastic increase in $K$. A detailed study of the effects of temperature on the reduction of GO found that O reduction and partial exfoliation of graphitic layers starts at temperatures as low as 127°C [29]. The exfoliation accompanies partial reduction as a consequence of the substantial loss of the oxygen surface groups. When GO is treated at higher temperatures, $T \geq 600$°C, the reduction improves, with a loss of O and H and a conversion of hybridized carbon atoms from $sp^3$ into $sp^2$ [29]. The authors of Ref. [29] also noted that $T=1000$°C appears to be a critical temperature in GO treatment from the point of view of the efficiency of the reduction process, as the resulting graphene-like material contained <2% oxygen and 81.5% C $sp^2$ [29]. Our findings are in line with this report in terms of the loss of O, increased $sp^2$ content and enlargement of the graphene-like domains within each layer. All these factors together lead to the observed increase in the in-plane thermal conductivity and unusually high $K/K_\perp$ ratio.

It is interesting to note that despite a significant increase of the thermal conductivity for the thermally treated rGO film ($K$=61 W/mK at RT for rGO annealed at $T$=1000°C) it is still much lower than that in graphite or graphene ($K$=2000 W/mK at RT for basal planes of graphite and can exceed this value in large graphene layers) [8, 12]. The reason for this is that the phonon thermal transport in rGO films is still limited not by intrinsic properties of graphene layers but by the grain and disorder scattering. The samples treated at $T$=1000°C only start showing the signs of the intrinsic Umklapp scattering. There have been only a few theoretical computational studies reported for thermal conductivity of GO and rGO [49-50]. For this reason there is a need in the experimentally validated theoretical models that can be used for optimization of the thermal transport in rGO films.

We now turn to theoretical interpretation of the experimental results. For the theoretical investigation of the thermal conductivity in rGO we adopted our approach, developed for thermal conductivity of graphite in Ref. [11]. In our simulation we employ graphite phonon energy dispersions, obtained within Born – von Karman model of lattice dynamics [51-52], and consider oxygen and other impurities as point defects. Following the approach of Ref. [11] we treat the phonon transport in rGO as two-dimensional (2D) for phonons with frequencies $\omega_s > \omega_{c,s}$ and three-dimensional (3D) for phonons with $\omega_s \leq \omega_{c,s}$, where $\omega_{c,s}$ is the low-bound cutoff frequency of the $s$-th phonon branch. The in-plane thermal conductivity $K^{\text{in-plane}}$ is given by

$$K^{\text{in-plane}} = K^{3D} + K^{2D},$$

$$K^{3D} \equiv \frac{\hbar^2}{4\pi^2 k_B T^2} \sum_s \int_0^{\omega_{c,s}} q_{z,s}(\omega)\omega^2 \tau_s(\omega) \upsilon_s^{\|}(q_{\|}) \frac{\exp(\hbar\omega/k_B T)}{[\exp(\hbar\omega/k_B T)-1]^2} q_{\|} d\omega, \quad (1)$$

$$K^{2D} = \frac{\hbar^2}{4\pi^2 k_B T^2} \sum_s \frac{\omega_{c,s}}{\upsilon_s^{\perp}} \int_{\omega_{c,s}}^{\omega_{\max,s}} \omega^2 \tau(\omega) \upsilon_s^{\|}(q_{\|}) \frac{\exp(\hbar\omega/k_B T)}{[\exp(\hbar\omega/k_B T)-1]^2} q_{\|} d\omega.$$

In Eq. (1) $\tau_s(\omega)$ is the relaxation time for a phonon with the frequency $\omega$ from the $s$-th acoustic phonon branch, $\vec{q}(q_{\|}, q_z)$ is the phonon wave vector, $\upsilon_s = d\omega/dq^{\|}$ is the in-plane phonon group velocity for $s$-th branch, $T$ is the temperature, $k_B$ is the Boltzmann's constant, $\hbar$ is the Planck's constant. The summation in Eq. (1) is performed over six lowest phonon branches: in-plane longitudinal acoustic $LA_1$, in-plane transverse acoustic $TA_1$, out-of-plane transverse acoustic $ZA$, in-plane longitudinal acoustic-like $LA_2$, in-plane transverse acoustic-like $TA_2$ and out-of-plane transverse acoustic-like $ZO'$.

We assume that the Umklapp scattering ($U$), point-defect scattering ($PD$) and scattering on ordered clusters edges ($E$) are the main mechanisms limiting the thermal conductivity in rGO. The total phonon relaxation time $\tau$ was calculated using the Matthiessen's rule as [11-12, 3, 53-55]: $1/\tau_s = 1/\tau_{PD,s} + 1/\tau_{U,s} + 1/\tau_{E,s}$, where $\tau_{U,s}(\omega) = M\upsilon_s^2 \omega_{\max,s}/(\gamma_s^2 k_B T[\omega]^2)$, $\tau_E(\omega) = L/\upsilon_s^{\|}$ and $\tau_{PD,s}(\omega) = 4v_s^{\parallel} / (S_0 \Gamma q^{\parallel} \omega^2)$ . Here $\gamma_{LA_1,LA_2} = 2$ , $\gamma_{TA_1,TA_2} = 1$ and $\gamma_{ZA,ZO'} = -1.5$ is the branch-dependent average Gruneisen parameters, $\omega_{max,s}$ is the maximum frequency of $s$-th phonon branch, $S_0$ is the cross-section area per atom, $M$ is the graphite unit cell mass, $\Gamma$ is the measure of the strength of the point-defect scattering due to mass-difference and $L$ is the average length of ordered sp$^2$ or sp$^3$ clusters. The values of $\omega_{c,s}$ were determined from the phonon spectra as the highest energy of $s$-th branch along c-axis direction: $\omega_{c,LA_1/LA_2} = 89$ cm$^{-1}$, $\omega_{c,TA_1/TA_2} = 89$ cm$^{-1}$ and $\omega_{c,ZA/ZO'} = 32$ cm$^{-1}$. The strength of the point-defect scattering $\Gamma$ was estimated from the following formula [3]: $\Gamma = \sum_i c_i (\Delta M_i / M_C)^2$ , where $\Delta M_i = M_{d,i} - M_C$ is the difference between mass of the point-defect $M_{d,i}$ and carbon mass $M_C$, $c_i$ is the ratio between the concentrations of the defects $i$ and carbon atoms. We consider an impurity atom attached to a carbon atom as a point-defect in the graphite lattice. Our XPS study revealed that the three main impurity atoms present in our samples are O, N, and S. The $\Gamma$ parameter was calculated for each sample separately taking into account actual concentrations of defects. The defect concentrations and obtained values of $\Gamma$ are listed in Table III.

**Table III:** Elemental Composition and $\Gamma$ Parameter Values

|  | C (%) | O (%) | S (%) | N (%) | $\Gamma$ |
|---|---|---|---|---|---|
| **GO** | 65.9 | 29.2 | 3.7 | 1.1 | 1.208 |
| **rGO (T=300$^o$C)** | 89.4 | 10.3 | 0.3 | 0 | 0.229 |
| **rGO (T=600$^o$C)** | 90.6 | 8.6 | 0.4 | 0.4 | 0.206 |
| **rGO (T=1000$^o$C)** | 91.9 | 6.7 | 1.0 | 0.4 | 0.213 |

The dependence of the in-plane thermal conductivity, calculated from Eq. (1) is shown in Figure 6 (a) as a function of the average length $L$ of the ordered graphitic clusters. The results are presented for different values of $\Gamma$. The thermal treatment enhances the thermal conductivity of rGO due to the following reasons: (i) decrease of the defect concentrations, (ii) increase of the lateral dimensions of the ordered clusters and (iii) the rise of the sp$^2$ fraction. The oxygen concentration permanently decreases with temperature in our samples (see Table III). Nevertheless, the parameter $\Gamma \sim 0.21$ is estimated for all thermally treated rGO due to difference in $S$ and $N$ content. It means that the point-defect scattering is roughly the same for all treated rGO and by a factor of six weaker than in GO. The latter allows one to make a conclusion that increasing of the lateral dimensions of the ordered clusters is the key reason for the thermal conductivity enhancement in rGO samples. The restoring of the highly ordered graphene-like lattice after treatment was reported in Refs. [29, 34-35]. Comparing the theoretical and experimental thermal conductivity data we can roughly estimate the average length of the ordered clusters: it increases from ∼ 3.5 nm in GO to ∼ 500 nm in rGO (T=1000$^o$C). The obtained $L$ ∼ 500 nm are in range of the average grain lengths 250 nm – 30 μm reported for polycrystalline graphene [56] and graphite [57]. The actual average cluster length in our rGO samples could be even lager due to possible additional phonon scattering on vacancies and dislocations. The vacancy concentration ∼ 0.5% increases theoretical $L$ up to 800 nm for rGO (T=1000$^o$C). The increase of $L$ in rGO accelerates with temperature: in rGO (T=600$^o$C) the $L$ value is by a factor of three larger than in rGO (T=300$^o$C) while in rGO (T=1000$^o$C) it increases by a factor of 16 as compared with rGO (T=600$^o$C). The thermal conductivity of rGO can be increased up to ∼ 500 W/mK for samples with larger grains and reduced impurities. In Figure 6 (b) we illustrate the impact of the oxygen reduction on the thermal conductivity of rGO. For this plot we assumed that only oxygen impurities are present in rGO. Decreasing O concentration from 50% to 1% increases $K$ by a factor of 4 – 24 depending on the average cluster size $L$. Our calculations show that S and N impurities in rGO (T=1000$^o$C) suppress $K$ by 28%. Removing these impurities together with reducing O concentration down to 1% allows to obtain $K$ ∼ 300 W/mK at $L$ ∼ 500 nm.

The decrease of the cross-of-plane thermal conductivity in rGO with the higher treatment temperature can be qualitatively explained by an increase in the number and size of the "air pockets" between rGO multilayers (see inset to Figure 6 (a) and Figures 1 (b-d)). The average thickness of rGO film treated at 1000$^0$C is by a factor of four larger than that of the GO film indicating the increase in the "air pocket" volume. The cross-plane thermal transport is affected by the "air pockets" much more than the in-plane thermal transport. In the Maxwell-Garnett's effective medium approximation, the cross-plane thermal conductivity can be estimated as [58]:

$$K_{rGO}^{\perp}(\varphi) = K_{GO}^{\perp} \frac{(1-\varphi)(K_{air} + 2K_{GO}^{\perp}) + 3\varphi K_{air}}{(1-\varphi)(K_{air} + 2K_{GO}^{\perp}) + 3\varphi K_{GO}^{\perp}}, \qquad (2)$$

where $\varphi$ is the volume fraction of the "air pockets", $K_{GO}^{\perp}$ is the thermal conductivity of the non-treated GO and $K_{air}$ is the thermal conductivity of air. Using $K_{GO}^{\perp} \approx 0.18\,\text{W/mK}$, $K_{air} \approx 0.026\,\text{W/mK}$, and $\varphi \sim 0.5-0.6$ we obtained $K_{rGO}^{\perp}$ ~0.075-0.09 W/mK, which is in a good agreement with the experimental value of 0.09±0.01 W/mK for rGO (T=1000 $^{\circ}$C) at T = 20 $^{\circ}$C.

[Figure 6: Theory]

In conclusions, we investigated the thermal conductivity of free-standing rGO films subjected to a high-temperature treatment $T$=300$^{\circ}$C, 600$^{\circ}$C and 1000$^{\circ}$C. It was found that the high-temperature treatment dramatically increased the room-temperature in-plane thermal conductivity, $K$, from 2.94 W/mK in the reference GO film to 61.8 W/mK in the rGO film annealed at $T$=1000$^{\circ}$C. The cross-plane thermal conductivity, $K_{\perp}$, revealed an intriguing opposite trend of decreasing from ~0.18 W/mK in the reference GO film to ~0.09 W/mK in the rGO film annealed at $T$=1000$^{\circ}$C. The obtained films demonstrated an exceptionally strong anisotropy of the thermal conductivity, $K/K_{\perp}$ ~ 675, which is substantially larger even than in the high-quality graphite ($K/K_{\perp}$~100). The electrical resistivity of the annealed films reduced from a 0.5MΩ/□ range in the reference GO film to 1 Ω/□ − 19 Ω/□ in the high-temperature treated rGO films. The observed modifications of the in-plane and cross-plane thermal conductivity components resulting have been explained theoretically. The increase of the in-plane thermal conductivity is due to restoration of C sp$^2$ bonds, decreased phonon scattering on O and other impurities and increase in the sp$^2$ grains. The decrease of the cross-plane thermal conductivity after high-temperature annealing is due to appearance of "air pockets" and softening of the restoring forces in this direction. The strongly anisotropic heat conduction properties of rGO films treated at high temperature can be useful for applications in thermal management, which requires materials which can remove excess heat (high $K$) along one direction and shield from heat (lower $K_{\perp}$) along the perpendicular direction.

**METHODS**

**Sample Preparation and Annealing Procedure:** The graphene oxide was prepared by a modified Hummers method [59-61]. With the obtained graphene oxide slurry, a concentrated dispersion was prepared (10 mg/mL) and placed in a glass mold of the desired dimensions. The water was first evaporated at room temperature and atmospheric pressure and then at 60ºC in a vacuum oven overnight. The graphene oxide paper was finally detached from the mold. For the thermal treatments a Carbolite (GHA 12/450) tube furnace equipped with a quartz tube was used. The samples were placed in a graphite holder and then heated up (2ºC/min) up to the desired temperature (300ºC, 600ºC, 1000ºC) in a $N_2$ atmosphere (500 sccm, controlled by a MFC). The samples were treated at the target temperature for 60 minutes and then cooled down, in order to obtain the rGO films.

**Thermal Measurement Details:** The measurements of cross-plane, $K_\perp$, and in-plane, $K$, thermal conductivity were performed using LFT (Netzsch LFA 477). The method is compliant with the international standards ASTM E-1461, DIM E-821 and DIN-30905. To perform LFT measurement, each sample was placed into a special stage and sample holder that fitted its size. The bottom of the stage was illuminated by a xenon lamp (wavelength $\lambda$=150 – 2000 nm) with millisecond energy pulses. The temperature of the opposite surface of the sample was monitored with a cryogenically cooled InSb infrared detector [62-63]. The design of the in-plane sample holder ensured that heat traveled ~5 mm inside rGO film along its plane, which is a much larger distance than its thickness, and thus, ensuring the in-plane values for thermal diffusivity $\alpha$. The specific heat, $C_p$, was measured independently with the same instrument using the graphite reference (for graphite: $C_p$=0.6364 J/gK at $^o$C, gradually increasing to 0.9799 J/gK at 125$^o$C). The cross-plane thermal conductivity, $K_\perp$, was determined from the equation $K_\perp=(m/V)\alpha_\perp C_p$, where thermal diffusivity, $\alpha_\perp$, was measured in the standard cross-plane configuration, $m$ is the mass and $V$ is the volume. Our XPS study shows that temperature treatment decreases the sample mass due to removing oxygen, nitrogen and sulfur impurities:

$$m = \gamma m_0, \ \gamma = \frac{(M_c+f^O M_O+f^S M_S+f^N M_N)}{(M_c+f_0^O M_O+f_0^S M_S+f_0^N M_N)}, \qquad (3)$$

where $m_0$ is the mass of non-treated sample, $M_i$ is the molar mass of the $i$-th element ($i$=C,O, S, N), $f_0^i$ and $f^i$ is the ratio between concentration of $i$-th element and carbon in non-treated and treated samples, respectively. The ratio $m/V$ we expressed through the ratio between thickness of treated ($H$) and non-treated ($H_0$) samples as follows: $m/V = \gamma \rho_0 H_0/H$, where $\rho_0$ is the true density measured independently. Finally cross-plane thermal conductivity is given by

$$K_\perp = \gamma \rho_0 (H_0/H) \alpha_\perp C_{p.} \qquad (4)$$

The value of $\alpha_\perp$ was used as an input parameter during the in-plane measurements of the thermal diffusivity $\alpha$. An iterative scheme was used to separate the cross-plane (axial) $\alpha_\perp$ and in-plane (radial) diffusivities $\alpha$. Number of carbon layers, carrying in-plane heat is almost constant both for non-treated and treated samples, i.e. $H \approx H_0$. The small deviation between $H$ and $H_0$ is possible due to negligible change of carbon interlayer distance with treatment. The in-plane thermal conductivity was calculated from the Eq. (4), $K = \gamma \rho_0 \alpha C_p$, with the in-plane $\alpha$ value and $H = H_0$. To ensure accuracy, the instrument and data extraction procedures have been calibrated with materials of known thermal conductivity.

**XPS Measurement Procedures:** XPS characterization was carried out using a Kratos AXIS ULTRA XPS system equipped with an Al Kα monochrome X-ray source and a 165-mm mean radius electron energy hemispherical analyzer. The vacuum pressure was kept below $3 \times 10^{-9}$ Torr during the acquisition. The full spectrum and data acquisition parameters are presented in the *Supplementary Information*.

**Electrical Resistivity Measurements:** A micromanipulator probe station with a semiconductor analyzer to supply current and measure voltages were used to determine the sheet resistivity. Square samples were cut to approximately 0.5 mm$^2$ from GO and rGO films and probed with 25 μm radius tungsten pins. Four Ohmic contacts were established at each corner of a square sample for measurements of the resistance between the different pairs of contacts. The sheet resistance was calculated using the van der Pauw formula (see *Supplementary Information*).

## *Acknowledgements*

The work at UC Riverside was supported, in part, by the National Science Foundation (NSF) project CMMI-1404967 Collaborative Research Genetic Algorithm Driven Hybrid Computational Experimental Engineering of Defects in Designer Materials, NSF project ECCS-1307671 Two-Dimensional Performance with Three-Dimensional Capacity: Engineering the Thermal Properties of Graphene, and by STARnet Center for Function Accelerated nanoMaterial Engineering (FAME) – Semiconductor Research Corporation (SRC) program sponsored by The Microelectronics Advanced Research Corporation (MARCO) and the Defense Advanced Research Project Agency (DARPA). The XPS instrument used in this study was acquired with the funds from NSF under their Major Research Instrumentation Program DMR-0958796. AIC and DLN acknowledge financial support from the Moldova State grant 15.817.02.29F and STCU project 5937.

*Author Contributions*

A.A.B. conceived the idea, coordinated the project, led the thermal data analysis, contributed to the theory and wrote the manuscript; JDR and SR performed "laser flash" thermal measurements, Raman and XPS characterization and contributed to manuscript preparation; HM conducted UV Raman and optothermal measurements; DLN and AIC carried out theoretical modeling and computational studies; BA, AC and AZ performed the graphene based material synthesis and thermal reduction experiments.

*Author Information*

The authors declare no competing financial interests. Correspondence and requests for materials should be addressed to (A.A.B.) balandin@ee.ucr.edu

***Supporting Information Available:*** The supporting information provides additional SEM, XPS and thermal data.

# FIGURE CAPTIONS

**Figure 1:** Scanning electron microscopy images of the top view (a-c) and cross-sectional view (b-d) of representative free-standing films. The images (a-b) are of a reference GO film (untreated) and (c-d) are of rGO film (annealed at 600$^o$C). The surface becomes corrugated due to the "air pockets" and results in an average thickness increase from $H$=40 μm of the reference GO film to $H$=170 μm in rGO film annealed at 1000°C.

**Figure 2:** X-ray photoelectron spectroscopy data of the carbon signatures for (a) reference GO film and (b) rGO film thermally treated to 600$^o$C. The main peaks observed at ~284.6 eV (red line), 286.8 eV (green line) and 288.1 eV (blue line) correspond to sp$^2$ and sp$^3$ C, single bonded carbon – oxygen (C-O), and double-bonded carbon – oxygen (C=O), respectively. The expulsion of O by thermal treatment leads to the concentration of C exceeding 90% after 600°C annealing.

**Figure 3:** Raman spectra of rGO films under (a) visible (λ= 488 nm) and (b) UV (λ= 325 nm) laser excitations. The Raman spectra are shown (a) for rGO samples that underwent thermal treatment at 300$^o$C, 600$^o$C, and 1000$^o$C. The peaks at ~1350 cm$^{-1}$ and 1580 cm$^{-1}$ correspond to the D and G peaks, respectively, which have more pronounced separation with thermal treatment at 1000$^o$C. The 2D band and S3 peak also become well defined indicating that the films are moving from an amorphous state to a more ordered material. The ratio of the intensity of D peak to that of G peak in Raman spectra under UV excitation (b) suggests the reduction in defects and sp$^3$ bonds.

**Figure 4:** Experimental cross-plane thermal conductivity, $K_\perp$, as a function of temperature for rGO annealed at different temperatures and a reference GO film. Note that $K_\perp$ is reduced for high temperature annealed samples as compared to that of reference GO film.

**Figure 5:** Experimental in-plane thermal conductivity, $K$, as a function of temperature for rGO films annealed at different temperatures and a reference GO film. The higher annealing temperature results in progressively higher $K$ values. The room-temperature thermal conductivity, $K$, increases from 2.9 W/mK for the reference GO film to 61 W/mK for the rGO film annealed at 1000°C.

**Figure 6:** Calculated in-plane thermal conductivity of GO and rGO at T = 20 C as a function of the average cluster size $L$. The data provides (a) comparison between the theoretical calculations and experimental data and (b) illustrates the effect of the oxygen impurity.

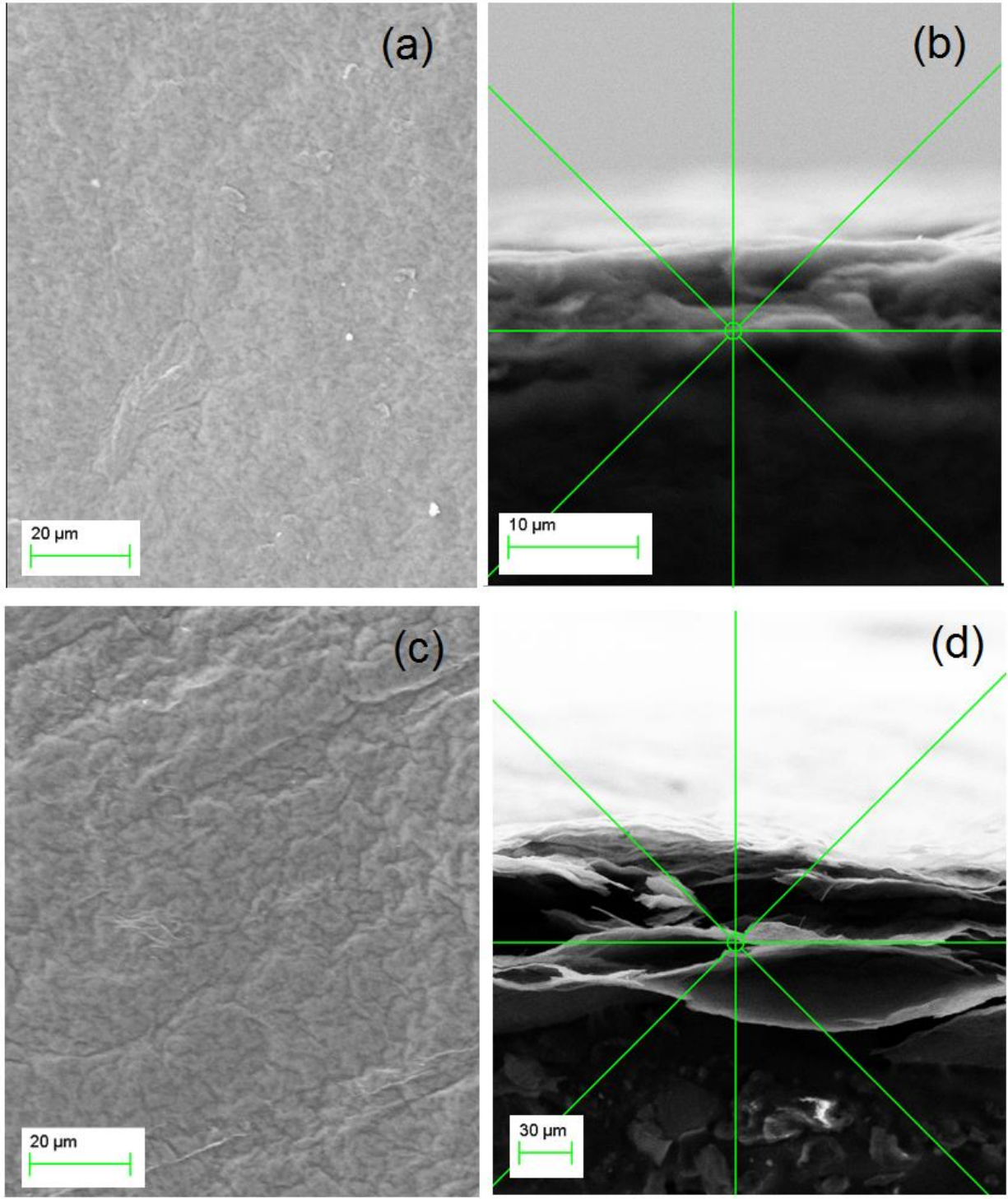

Figure 1 of 6: Renteria et al.

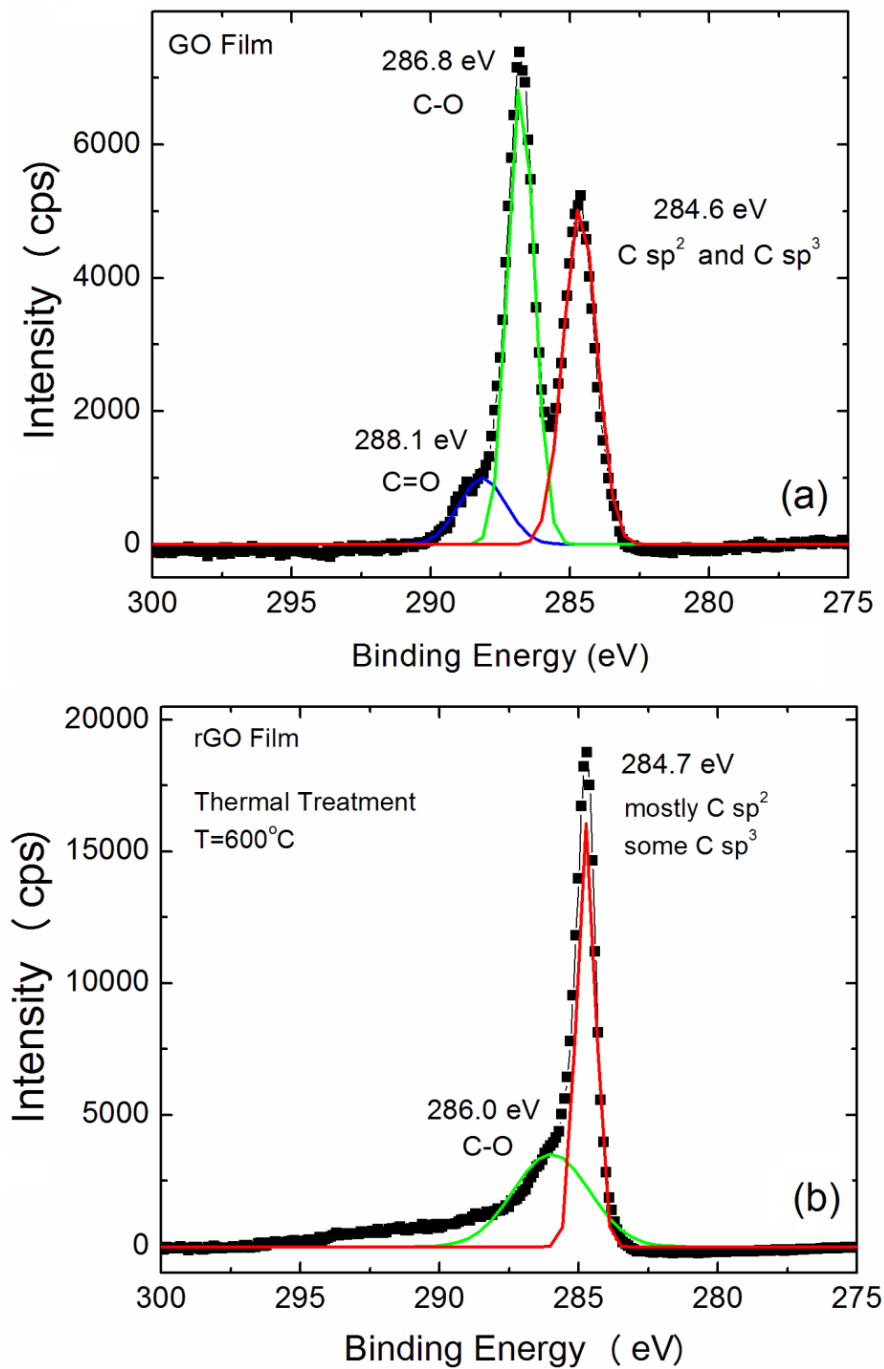

Figure 2 of 6: Renteria et al.

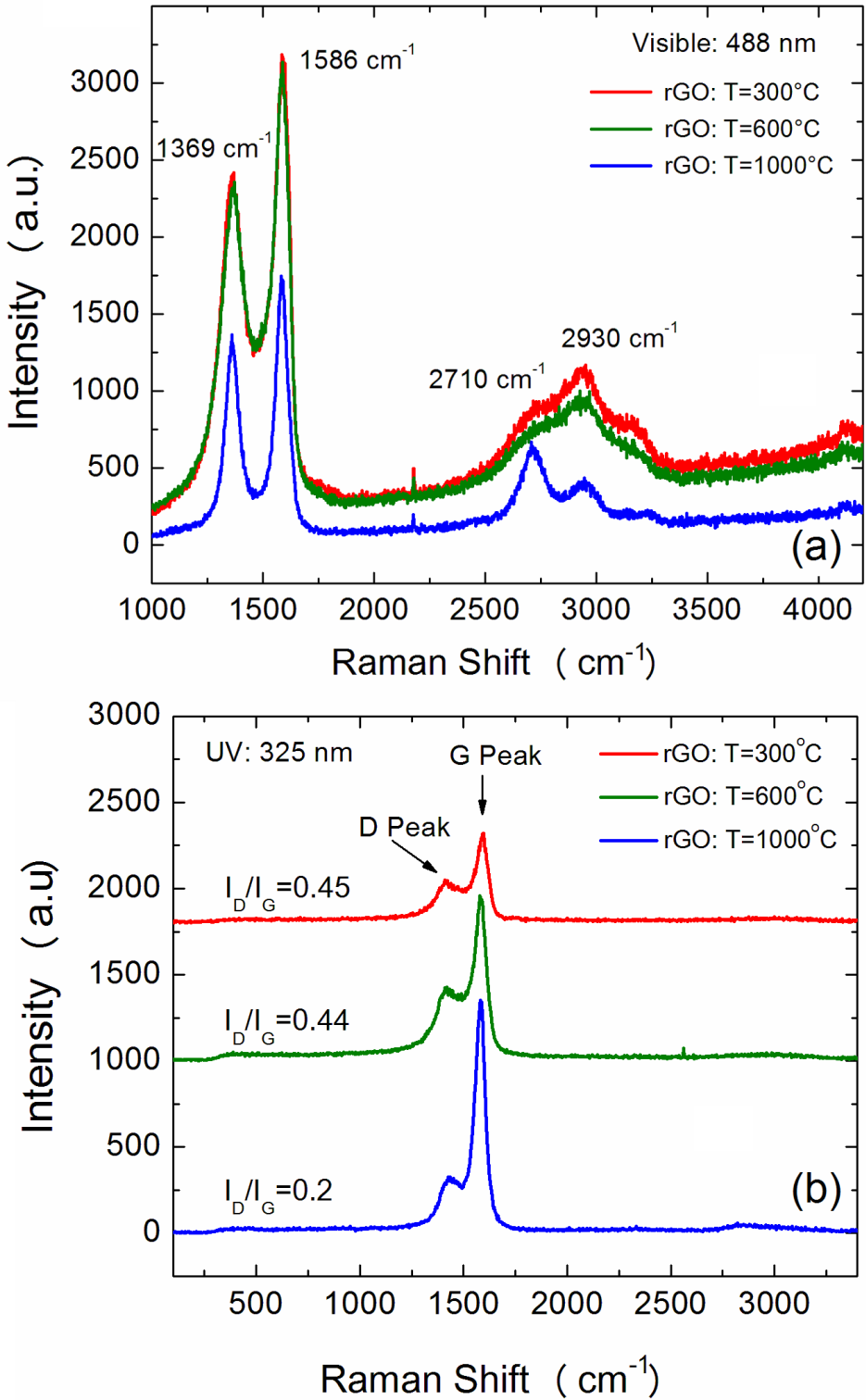

Figure 3 of 6: Renteria et al.

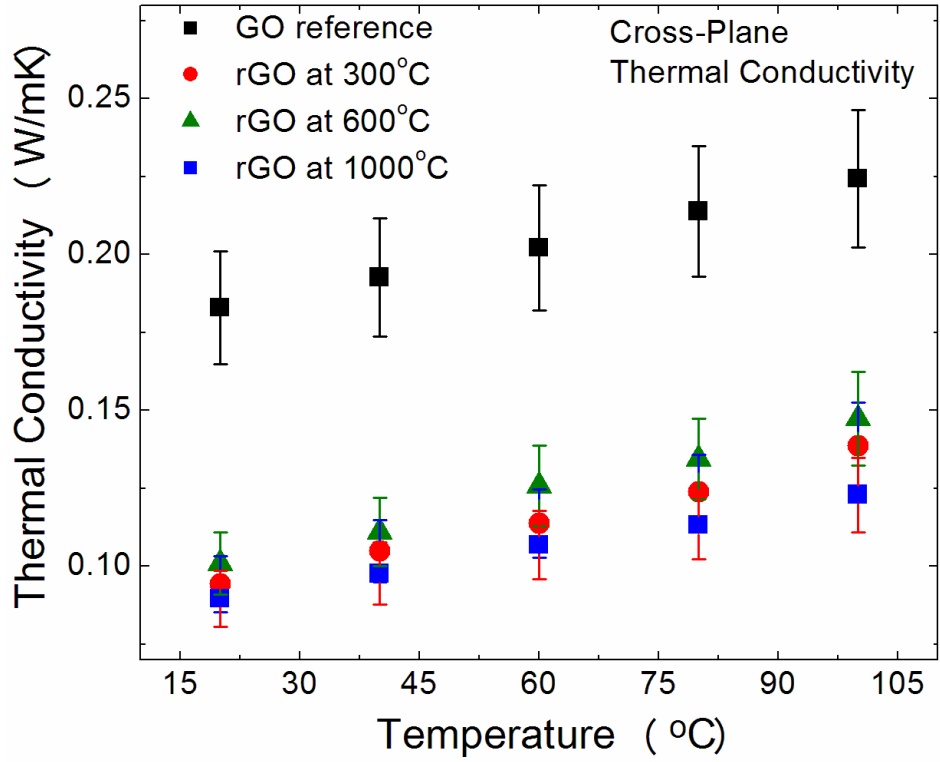

Figure 4 of 6: Renteria et al.

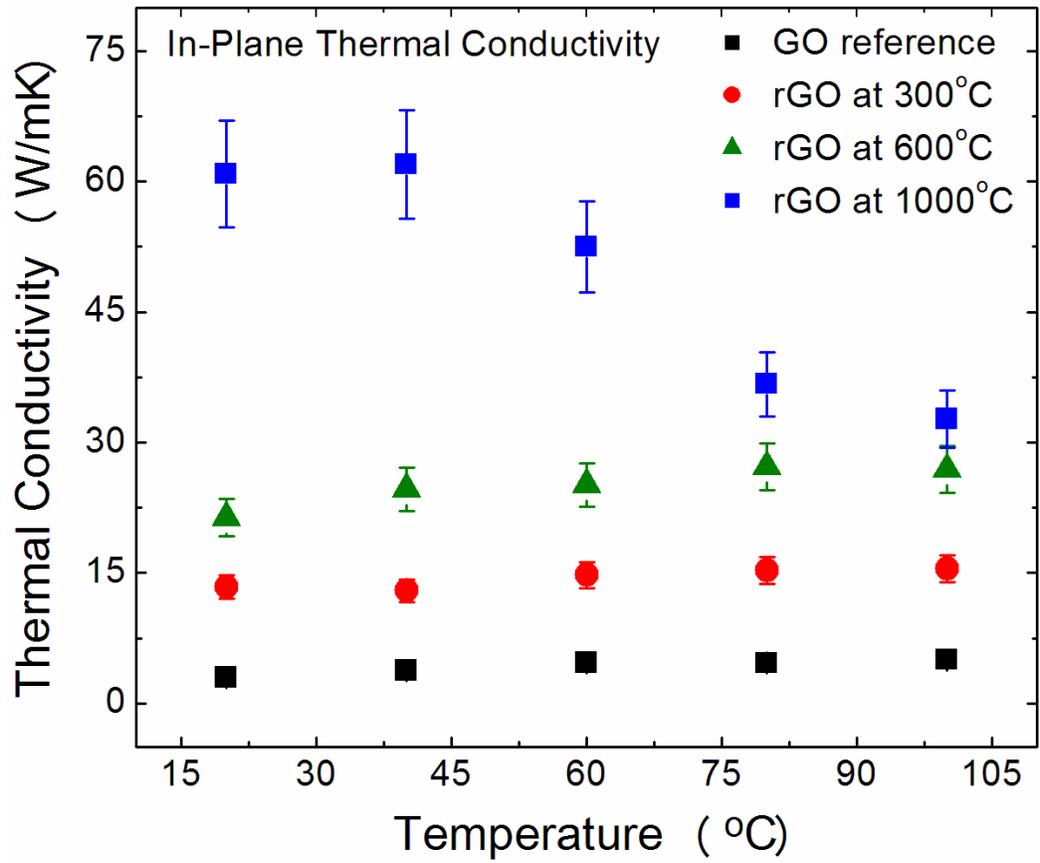

Figure 5 of 6: Renteria et al.

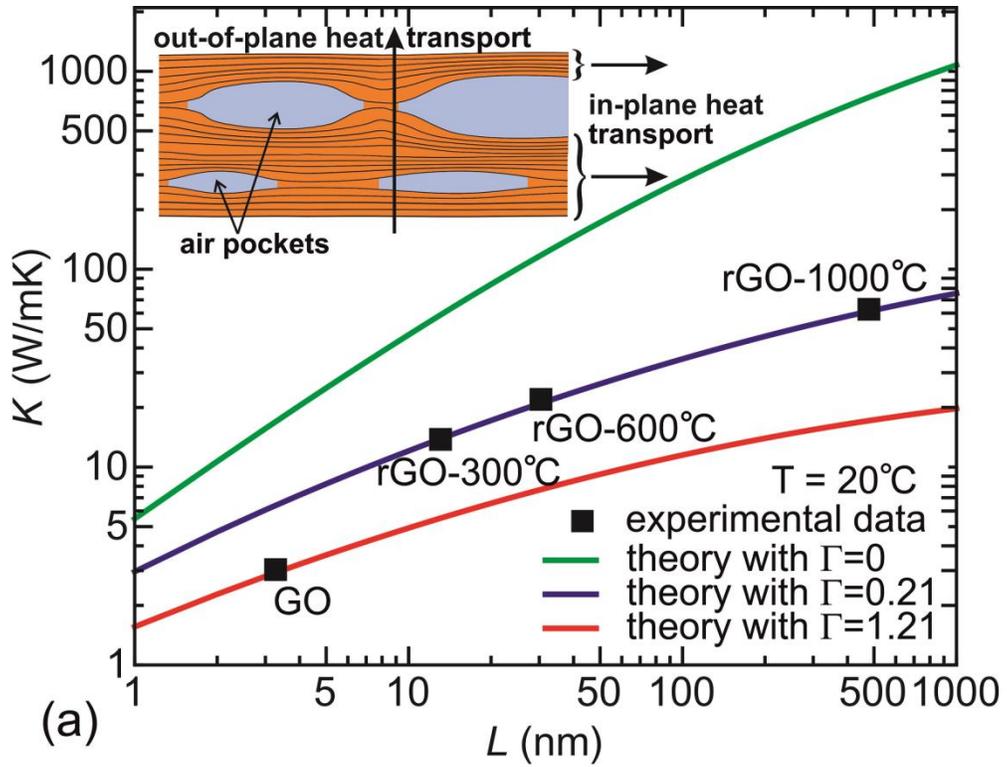

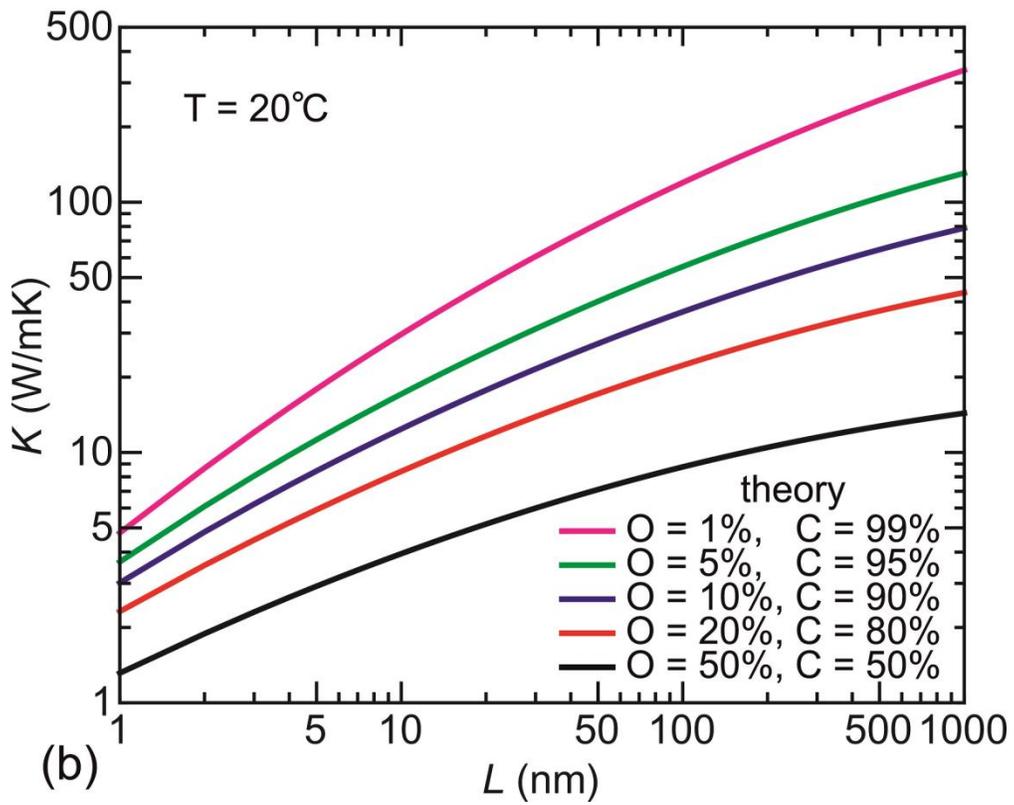

Figure 6 of 6: Renteria et al.